%% file: LMC_P3.tex
\documentclass{PoS}
\usepackage{graphicx}

\newcommand{\dgr}{\ensuremath{^\circ}}
\newcommand{\ergs}{erg\,s$^{-1}$}%
\newcommand{\cms}{cm$^{-2}$\,s$^{-1}$}%
\newcommand{\cmstev}{cm$^{-2}$\,s$^{-1}$\,TeV$^{-1}$}%

\newcommand{\dem}{DEM~L241}
\newcommand{\cxou}{CXOU~053600.0$-$673507}
\newcommand{\nb}{N\,157B}
\newcommand{\psr}{PSR\,J0537$-$6910}
\newcommand{\CAL}{CAL~60}
\newcommand{\hessj}{HESS~J0536$-$675}
\newcommand{\pthree}{LMC~P3}

\newcommand{\ls}{LS~5039}
\newcommand{\jten}{1FGL~J1018.6$-$5856}
\newcommand{\psrb}{PSR~B1259$-$63/LS~2883}

\newcommand{\fermi}{$Fermi$-LAT}
\newcommand{\hess}{H.E.S.S.}
\newcommand{\xmm}{{\it XMM-Newton}}
\newcommand{\chandra}{{\it Chandra}}

\newcommand{\Lint}{$(1.4 \pm 0.2) \times 10^{35}$\,\ergs}
\newcommand{\Lon}{$(5 \pm 1) \times 10^{35}$\,\ergs}

\newcommand{\SigInt}{$6.4\,\sigma$}
\newcommand{\SigOn}{$7.1\,\sigma$}

\def\gr{$\gamma$-ray}
\def\grs{$\gamma$ rays}

\include{includes}

\title{Discovery of VHE Gamma-Ray Emission from the Binary System LMC P3}

\ShortTitle{Discovery of VHE Gamma-Ray Emission from the Binary System LMC P3}

\author{\speaker{Nukri Komin}\\
        School of Physics, University of the Witwatersrand, Johannesburg, South Africa.\\
        E-mail: \email{nukri.komin@wits.ac.za}}

\author{Maria Haupt\\
		DESY, Zeuthen, Germany.
		}

\author{for the H.E.S.S. Collaboration \thanks{https://www.mpi-hd.mpg.de/hfm/HESS/pages/collaboration/}\\
}

\abstract{Recently, the \gr\ emission at MeV and GeV energies from the object \pthree\ in the Large Magellanic Cloud has been discovered to be modulated with a 10.3-days period, making it the first extra-galactic \gr\ binary. This work aims at the detection of TeV \gr\ and the search for modulation of the signal with the orbital period of the binary system. The \hess\ data set has been folded with the known orbital period of the system in order to test for variability of the emission. 
    Energy spectra are obtained for the orbit-averaged data set and for orbital phases in which the TeV flux is found at its maximum. TeV \gr\ emission is detected with a statistical significance of \SigInt. The data clearly show variability which is phase-locked to the orbital period of the system. Periodicity cannot be deduced from the \hess\ data set alone. The orbit-averaged luminosity in the $1-10$\,TeV energy range is \Lint. A luminosity of \Lon\ is reached during 20\% of the orbit, when the MeV/GeV emission is at its minimum.
}

\FullConference{35th International Cosmic Ray Conference --- ICRC2017\\
		10--20 July, 2017\\
		Bexco, Busan, Korea}

\begin{document}

\section{Introduction}

More than 60\% of all stellar systems containing high-mass stars (spectral type B2 or earlier) are binary or multiple systems \cite{2013ARA&A..51..269D}. When the more massive of the stars in these systems ends its life in a supernova explosion, a binary system is left behind where a compact object, either a neutron star or a black hole, is orbiting the remaining star.
Matter accretion from the star to the compact object may lead to X-ray emission. These objects are known as X-ray binaries. If the companion is also a massive star, the system is called a high-mass X-ray binary (HMXB).
When the emission of these objects dominates in the \gr\ regime (high-energy (HE) \grs : $0.1 - 100$\,GeV, very-high-energy (VHE) \grs : $> 100$\,GeV) then they are called \gr\ binaries. Up to now only six \gr\ binaries are known.
The \gr\ emission arises either from the interaction of the pulsar and stellar winds or from accretion onto a black hole or neutron star. The companion stars in these systems are either O-type or Be stars. The nature of the compact objects are generally unknown, with the exception of \psrb\ where the detection of pulsed emission \cite{1992ApJ...387L..37J} shows that the compact object is a neutron star. A review of \gr\ binaries and their properties is given by \cite{2013A&ARv..21...64D,Dubus}.

In order to identify previously undetected \gr\ binaries, \cite{Fermi} performed a blind search for periodic emission in the \fermi\ data. They found that the MeV/GeV \gr\ signal of \pthree\ is periodic with a period of $10.301 \pm 0.002$~days. These authors defined phase~0 of this system to correspond to the maximum in the gamma-ray emission at MJD $57410.25 \pm 0.34$. 
\pthree\ is a previously unidentified \gr\ source \cite{FermiLMC} located in the Large Magellanic Cloud (LMC).
The position of \pthree\ is consistent with the position of \CAL, which is a soft X-ray source discovered by \cite{1981ApJ...248..925L}. \cite{1985AJ.....90...43C} identify a star of spectral type O5~III(f) as likely counterpart of this X-ray source. Subsequent X-ray observations with \xmm\ \cite{Bamba} and \chandra\ \cite{Seward} confirm a point-like X-ray source which is named \cxou. \cite{Seward} already concluded from the variabilities of the X-ray flux and the radial velocity of Balmer absorption lines that this object is likely a binary system. \pthree\ is located in the supernova remnant \dem, making it the third X-ray binary found in an observable supernova remnant after SS433/W\,50 \cite{SS433} and SXP\,1062 \cite{SXP1062}.

Little is known of the orbital parameters of the system. The most precise measurement of the orbital period comes from the HE \gr\ emission. \cite{Fermi} also analysed the radial velocity of the star and found an orbital period of 10.1~days and a superior conjunction of the companion star at MJD $57408.61 \pm 0.28$. The mass function prefers a neutron star as compact object for a wide range of inclinations, but a black hole cannot be ruled out. The X-ray and radio emission of this object is modulated with the 10.3-days period, but is out of phase with the \gr\ emission \cite{Fermi}.

Showing periodic \gr\ emission from a system with an O5 companion star makes this object clearly a high-mass \gr\ binary. It is the first such object discovered outside the Milky Way. With a \gr\ luminosity in the energy range from 200~MeV to 100~GeV of $2.5\times10^{36}\,$\ergs\ \cite{FermiLMC} it is also the most luminous \gr\ binary known so far.

   \begin{figure}
   \centering
   \includegraphics[width=0.7\hsize]{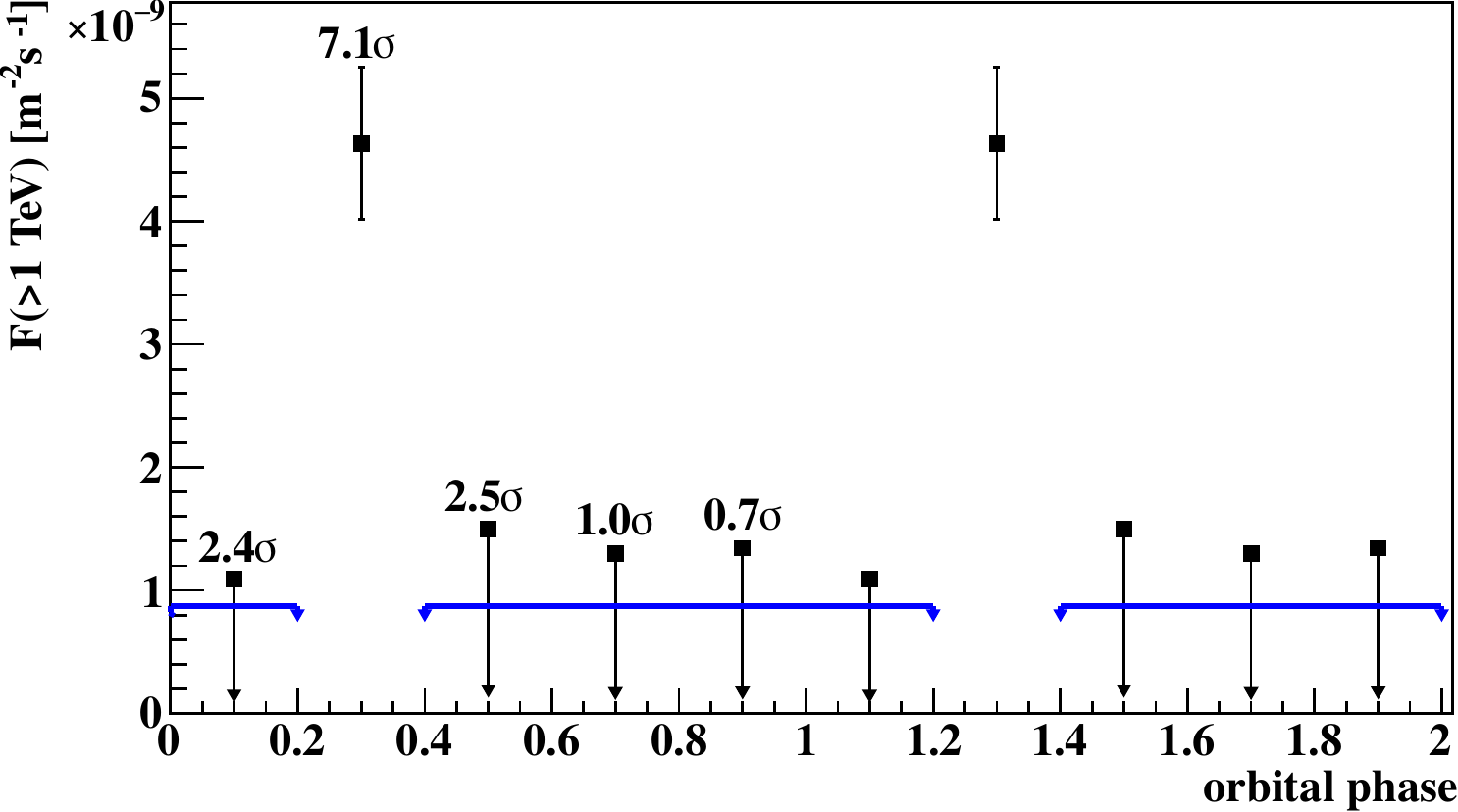}
      \caption{Folded TeV \gr\ light curve. Orbital phase 0 is defined at the maximum of the \fermi\ emission at MJD 57410.25. For better readability two orbits are shown. The flux is computed in orbital phase bins of equal size and similar exposure. Error bars represent $1\,\sigma$ statistical uncertainty. For the phase bins without significant detection upper limits (at 95\% confidence level) are given. The labels at the data points indicate the statistical significance of the excess in each phase bin. The blue lines denote the upper limit (at 95\% confidence level) on the flux in the off-peak region (orbital phase 0.4 to 0.2).}
         \label{fig:lightcurve}
   \end{figure}

   \begin{figure*}
   \centering
   \includegraphics[width=\hsize]{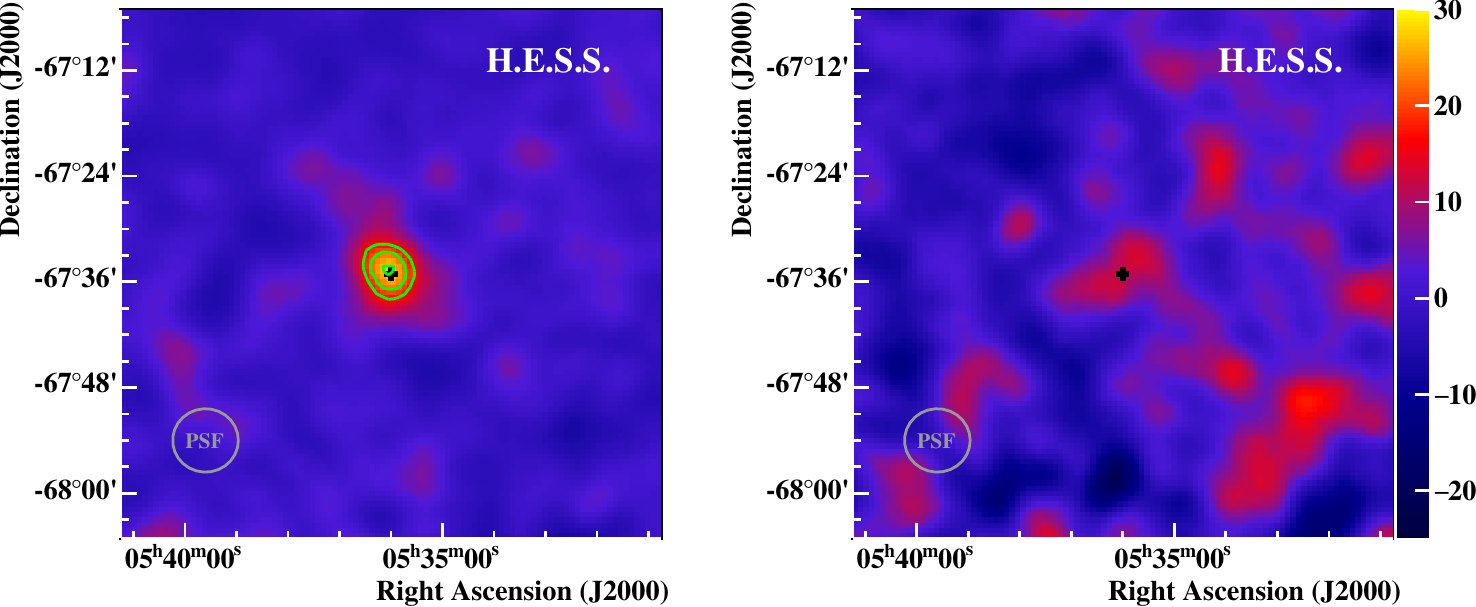}
   \caption{\hess\ excess maps for the on-peak (left panel) and off-peak (right panel) regions of the orbit. The excess is smoothed with the point-spread function of the instrument (68\% containment radius of $0.06\deg$, indicated by the grey circle)
. The cross indicates the test position. Both plots have the same range (colour bar on the right-hand side). The overlaid green contours represent 4, 5 and 6\,$\sigma$ statistical significance. The on-peak region covers orbital phases from 0.2 to 0.4, the off-peak region the orbital phases from 0.4 to 0.2. Orbital phase 0 is defined at the maximum of the \fermi\ emission at MJD 57410.25.}
              \label{fig:skymap}%
    \end{figure*}

\section{\hess\ Observations and Results}

The LMC has been observed extensively with the High Energy Stereoscopic System (\hess) since 2004. These observations lead to the discovery of three individual TeV \gr\ emitting sources \cite{N157B,LMCsurvey}. 
\hess\ is a system of five Imaging Cherenkov Telescopes, located in the Khomas Highland of Namibia at an altitude of 1800\,m. Located in the southern hemisphere it is currently the only instrument for TeV observations of the LMC. Phase I of \hess\ is sensitive to \grs\ from energies above 100 \,GeV up to several tens of TeV. The arrival direction of individual \grs\ can be reconstructed with an angular resolution of better than $0.1\dgr$, and their energy is estimated with an relative uncertainty of 15\%.
The data presented here were taken between 2004 and beginning of 2016 and add up to a total observation time of 277\,h, almost 70\,h more than what was used in the previous publication \cite{LMCsurvey}. The earlier observations were centred on the region around \nb\ resulting in relatively large camera offsets for \pthree. In the last years the observation strategy has been optimised in order to obtain a better exposure on \pthree. After correcting for dead time and camera offset angles the effective exposure time is 100\,h. 
In some of the observation runs the large \hess-II telescope was present. The data recorded with this telescope are ignored in the analysis in order to obtain a homogeneous data set.
The data were analysed using \textit{Model analysis} with high-resolution cuts \cite{Mathieu}, where the camera images are compared with simulations using a log-likelihood minimisation. The background was estimated from rings around each sky position to generate the gamma-ray image \cite{BGmodels} and from test regions with similar offsets from the camera centre for the spectral analysis \cite{BGmodels}. The energy threshold for this data set is 714\,GeV.

At the nominal position of \cxou\ an excess of 76.3 \grs\ is detected with a statistical significance of \SigInt\ (Table~\ref{tab:results}). The nightly light curve of the emission does not show any sign of variability. The search for periodic emission using a Lomb-Scargle test \cite{Lomb,Scargle} and the Z-Transformed Discrete Correlation Function \cite{ZDCF} does not show any sign of periodicity.
The sensitivity of \hess\ does not allow a detection of the object on a nightly basis, which is preventing the detection of variability and periodicity in the nightly light curve. 
Figure~\ref{fig:lightcurve} shows the light curve folded with the orbital period of the system of 10.301~days, where orbital phase 0 is defined at the maximum of the \fermi\ light curve at MJD $57410.25$ \cite{Fermi}. Significant emission is detected only at an orbital phase between 0.2 and 0.4 with a pre-trial significance of \SigOn\ (corresponding to $6.9\,\sigma$ post-trial after 5 trials). All other phase bins do not show any emission (significances less than $2.5\,\sigma$, see Fig.~\ref{fig:lightcurve}). 
All phase bins have roughly the same exposure (between 18 and 21 hours).
The fit of a constant to the flux values in the orbital phase bins results in a $\chi^2$ value of 27.03 for 4 degrees of freedom. The $\chi^2$ probability that the folded light curve is constant is less than $1.95 \times 10^{-5}$.
The emission is clearly variable and it is phase-locked to the orbital period of the system. Therefore, the detected TeV \gr\ emission can be associated to the binary system \pthree.
  

   \begin{table*}
      \caption[]{Statistical results and spectral parameters for different orbital phase bins of \hessj. The on-peak region covers orbital phases from 0.2 to 0.4, the off-peak region the orbital phases from 0.4 to 0.2. Orbital phase 0 is defined at the maximum of the \fermi\ emission at MJD 57410.25.}
         \label{tab:results}
\begin{tabular}{lcccccc}
\hline
\hline
orbital phase bin 	& $\Phi_{1\,\mathrm{TeV}}$ 
					& $ \Gamma$
					& $F (> 1\,\mathrm{TeV})$
					& $L (1 - 10\,\mathrm{TeV}, 50\,\mathrm{kpc})$ 
					\\
			 		& $[10^{-13}$ \cmstev$]$ 
					& 
					& $[10^{-13}$ \cms$]$
					& $[10^{35}$ \ergs$]$ 
					\\
\hline
full orbit	& $2.0 \pm 0.4$
			& $2.5 \pm 0.2$	
			& $1.4 \pm 0.4$
			& $1.4 \pm 0.2$
			\\
on peak & $5 \pm 1$
			& $2.1 \pm 0.2$
			& $5 \pm 2$
			& $5 \pm 1$  
			\\
off peak	& -
			& $2.4$ (fixed)	
			& $<1.2$ (95\% CL)
			& $<1.2$ (95\% CL)
			\\
\hline
\hline
\end{tabular}
   \end{table*}

%
   \begin{figure}
   \centering
   \includegraphics[width=0.7\hsize]{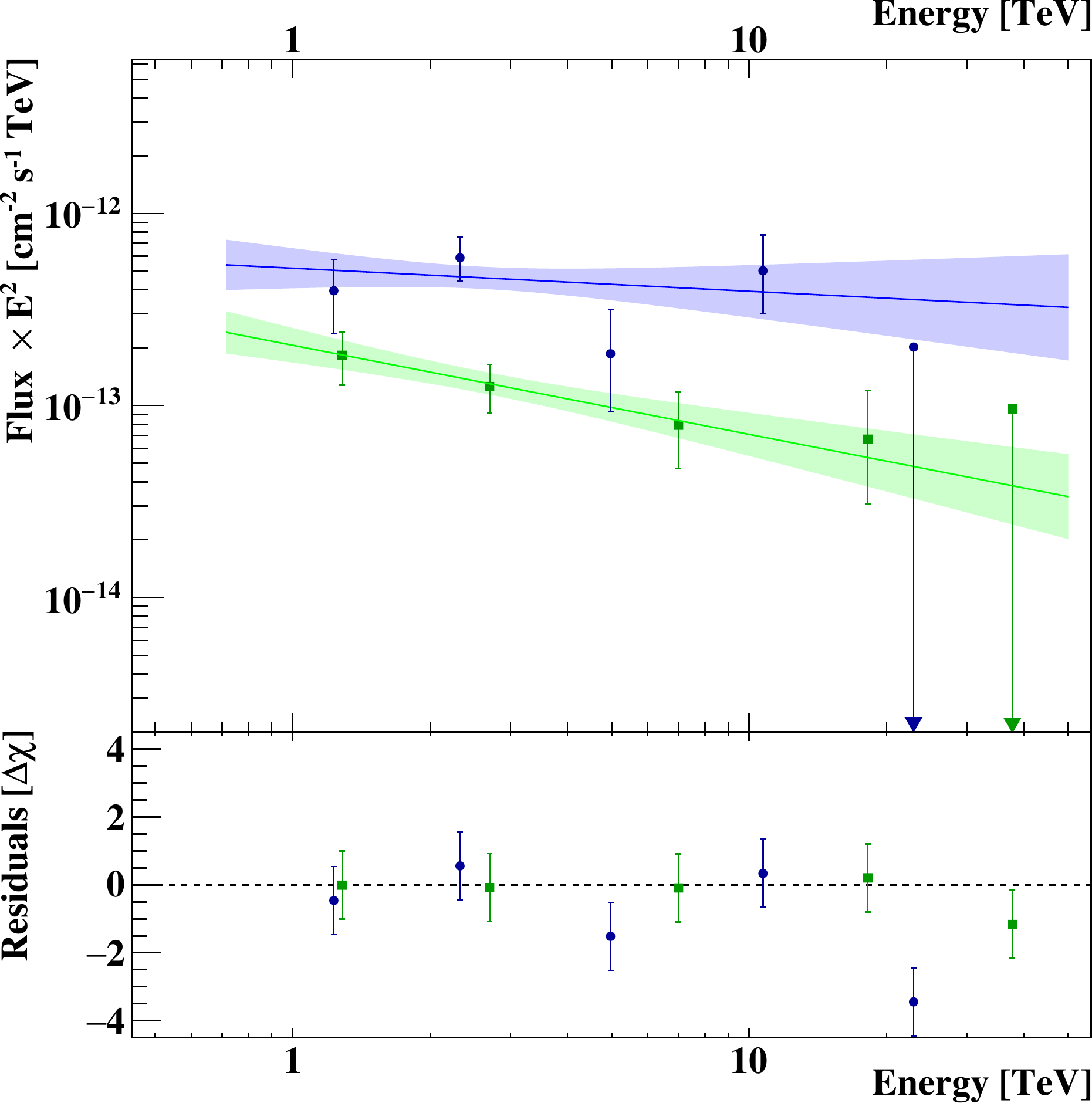}
      \caption{Spectral energy distribution averaged over the full orbit (green, square points) and for the on-peak orbital phase range (orbital phase from 0.2 to 0.4: blue, circle points). The data points have $1\,\sigma$ statistical error bars, upper limit are for a 95\% confidence level. The best fit and its uncertainty are represented by the solid lines and shaded areas, respectively.
              }
         \label{fig:spectrum}
   \end{figure}


Figure~\ref{fig:spectrum} shows the energy spectra for the full orbit and the on-peak part of the orbit. The spectra are fitted with a simple power law,
$\frac{dN}{dE} = \Phi_{1\,\mathrm{TeV}} \left( \frac{E}{1\,\mathrm{TeV}} \right)^{-\Gamma},$
and the best-fit parameters are summarised in Table~\ref{tab:results}. No high-energy cut-off of the spectra has been detected. For the off-peak region of the orbit an upper limit on the integrated \gr\ flux has been obtained. Most of the \gr\ emission is radiated during only 20\% of the binary's orbit, when it reaches a flux of about 4 times the orbit-averaged flux. 
\, 
\section{Discussion}

With an on-peak luminosity of \Lon\ \pthree\ is by far the most luminous \gr\ binary. Two main scenarios are usually invoked to explain the \gr\ emission from gamma-ray binaries: particle acceleration in a pulsar wind nebula (PWN) driven by the spin-down of the pulsar or accretion of the stellar wind on the compact object.

The \gr\ luminosity is about the same as the luminosity of the PWN \nb\ in the LMC. \nb\ is driven by the spin-down of its central pulsar \psr\ with a spin-down luminosity of $4.9 \times 10^{38}$\,\ergs. A putative pulsar in \pthree\ would need about the same spin-down luminosity. This would make it one of the four most luminous pulsars. That such a luminous pulsar remains undetected so far can be explained by absorption in the stellar photon field of the companion star.

In an alternative scenario stellar wind is accreted on the compact object. O-type stars have a mass loss through their stellar wind of at least $10^{-6}\,M_\odot\,\mathrm{year}^{-1}$, corresponding to $6 \times 10^{40}$\,\ergs. Therefore a fraction of about $10^{-5}$ of the acceretion power needs to be converted into \grs. It should be noted that in such a scenario the formation of jets may be expected (microquasars) which are usually detected at radio energies. No such jets are known for \pthree.

While the nature of the compact object and the mechanism that leads to the \gr\ emission are not yet clear, it can already be said that a very luminous pulsar or a strong stellar wind is necessary to provide the energy for the observed \gr\ emission.

In order to better understand the binary system optical observations to measure the orbital parameters are necessary. Furthermore, a deeper characterization of the system at VHEs will only come with the future Cherenkov Telescope Array (CTA), which will provide the required sensitivity to measure the relatively faint gamma-ray emission produced during the off-peak parts of the orbit.

\section{Summary}

Significant TeV \gr\ emission has been detected from the binary system \pthree\ with \hess  With an O-type companion star this system is similar to \ls\ and \jten. The relatively peaky light-curve is similar to \jten, but GeV and TeV emission are in anti-phase as is the case for \ls. The  on-peak luminosity  of \Lon\ makes it by far the most luminous \gr\ binary. A pulsar with a large spin-down luminosity of about $10^{38}$\,\ergs\ or a strong stellar wind are necessary to provide for the energy in \gr\ photons.

\end{document}

%% file: includes.tex
\def\aj{AJ}%
\def\araa{ARA\&A}%
\def\apj{ApJ}%
\def\apjl{ApJ}%
\def\apss{Ap\&SS}%
\def\aap{A\&A}%
\def\aapr{A\&A~Rev.}%
\def\mnras{MNRAS}%
\def\utw{\smash{\rlap{\lower5pt\hbox{$\sim$}}}}
\def\udtw{\smash{\rlap{\lower6pt\hbox{$\approx$}}}}

\def\sun{{\lower-2pt\hbox{$_\odot$}}}